\documentclass[aps,pre,twocolumn,reprint,tightenlines,amsmath,amssymb,amsfonts,showpacs,superscrpitaddress, floatfix]{revtex4-1}
\usepackage{graphicx}
\usepackage{dcolumn}
\usepackage{bm}
\usepackage{color}
\usepackage{csquotes}
\newcommand{\beq}{\begin{equation}}
\newcommand{\eeq}{\end{equation}}

\def\nothing#1{}
\begin{document}
\title{Shaking-induced crystallization of dense sphere packings}
\author{D. P. Shinde}
\email{shinde@bose.res.in}
\affiliation{Department of Theoretical Sciences, S. N. Bose National Centre for Basic Sciences, Calcutta, Calcutta 700098, India}
\author{Anita Mehta}
\email{anita@bose.res.in}
\affiliation{Department of Theoretical Sciences, S. N. Bose National Centre for Basic Sciences, Calcutta, Calcutta 700098, India}
\email{anita@bose.res.in}
\author{G. C. Barker}
\affiliation{Institute of Food Research, Norwich Research Park, Colney, Norwich,
NR4 7UA, United Kingdom }
\email{gary.barker@ifr.ac.uk}

\begin{abstract}
{We use a hybrid Monte Carlo algorithm to simulate the shaking of spheres at different vibrational amplitudes, and find that spontaneous crystallisation occurs in specific dynamical regimes. Several crystallising transitions are typically observed, leading to end states which can be fully or partially ordered, depending on the shaking amplitude, which we investigate using metrics of global and local order. At the lowest amplitudes, crystallisation is incomplete, at least for our times of observation. For amplitude ranges where crystallisation is complete, there is typically a competition between hexagonal close packed (hcp) or face-centered cubic (fcc) ordering. It is seen that fcc ordering typically predominates; in fact for an optimal range of amplitudes, spontaneous crystallisation into a pure fcc state is observed. An interesting feature is the breakdown of global order when there is juxtaposition of fully developed hcp and fcc order locally: we suggest that this is due to the interfaces between the different domains of order, which play the same role as dislocations.}. 
\end{abstract}


\maketitle

\section{\label{sec:INTRO}INTRODUCTION}
Hard sphere models are widely used in understanding the dynamics of thermal systems such as liquid-solid phase transitions \cite{Hansen}, nucleation and growth in colloids \cite{Gasser1,Blaaderen1,Frenkel} and glasses \cite{Tanaka}. The study of sphere packings in {\it athermal} systems received a boost when these became the centrepoint of models of dry granular media \cite{am}. While it is well known that hard spheres can sustain different degrees of packing, there has been little characterisation of either the associated structure or indeed the dynamical processes necessary to attain them. In this paper, we attempt to address some of these questions by examining the onset of spontaneous crystallisation in shaken granular assemblies. 

The lowest volume fraction at which an assembly of spheres is stable is known as the random loose packing limit, corresponding to a value of $0.55$ ($\phi_{rlp}$) \cite{Onoda}, while the highest value at which spheres can be packed in a fully disordered way is known as the random close packing limit ($\phi_{rcp}$) , corresponding to a value of $0.64$ \cite{Bernal,Scott1}. These numerical values have been the subject of experimental \cite{Mason,Aste,Blaaderen2}  and computational  \cite{Jodrey,Clarke} investigation, but are still widely regarded \cite {Torquato1} as approximate. 

At the other end of the spectrum, there is a conjecture by Kepler that that the maximum density of sphere packings is that of fcc structures, corresponding to a value of $0.74$ \cite{Hales}. What will concern us here are the spontaneous transitions from disorder to crystalline order that can occur in sphere packings; first observed in \cite{MehtaJCP}, these also have an analogue in the packings of ellipsoids \cite{Donev}. Since their theoretical prediction, such spontaneous transitions to crystallinity have been observed experimentally for sphere packings submitted to shear \cite{Tsai,Kudroli} or horizontal vibration \cite{Pouliquen}. The observed crystallinity can occur via fcc or hcp order, or indeed a mixture of the two. It has been suggested in the context of sheared colloidal suspensions  \cite{Gasser1,Frenkel,Gasser2,Cheng} that the fcc state is more stable than the hcp state \cite{Woodcock}; there is a similar observation also in the context of sheared granular spheres \cite{Kudroli}. One of the aims of this paper is to see whether this dominance of fcc ordering persists in the case spontaneous crystallisation of vertically vibrated granular packings, or if instead there is a coexistence of hcp and fcc in the asymptotic limit \cite{Makse}.

Another important question concerns the dynamical route to ordering, where it is well known that mechanical perturbations
such as shear and vibration can have distinct outcomes on granular configurations \cite{Nagel}. Accordingly we investigate
the kinds of ordering obtained as a function of the driving force, exploring both local and global features at different stages of cluster development in the packings generated by our computer simulations.

\section{\label{sec:SIMU}METHODS}

We use a three-dimensional Monte Carlo simulation algorithm \cite{MehtaPRL,MehtaPRE} to simulate the shaking of $N$ spheres. We briefly review the algorithm here before turning to its specific use in our current investigations.
Our simulations use monodisperse, hard spheres of unit diameter. The simulation cell is an open-topped box of size $10 \times 10 \times 10 $, and contains $N=1273$ spheres in all, with  periodic boundary conditions applied in the lateral directions. 
A unidirectional gravitational field acts downwards, i.e. along the negative z-direction. Initially the spheres are  placed in the cell using a sequential random close-packing procedure. The packing is then subject to a series of non-sequential, N-particle reorganisations. Each reorganisation is performed in three distinct parts: 
firstly a vertical expansion or dilation, secondly a Monte Carlo consolidation, and finally a non-sequential close packing procedure. We call each full reorganisation a shake cycle or, simply, a shake. The duration of our model shaking processes, and the lengths of other time intervals, are conveniently measured in units of the shake cycle.

The first part of the shake cycle is a uniform vertical expansion of the sphere packing, accompanied by random, horizontal, shifts of the sphere positions. Spheres are raised to new heights  and for each sphere, new lateral coordinates are 
assigned randomly, providing they do not lead to an overlapping sphere configuration. The (virtual) expansion 
introduces a free volume $A$ between the spheres and  facilitates their cooperative rearrangement during phases two and three of the shake cycle; $A$ is thus a measure of the amplitude of vibration. In the second phase of the cycle the whole system is compressed by a series of displacements of individual spheres. Spheres are chosen at random and displaced according to a hard sphere Monte Carlo algorithm. Finally, the sphere packing is stabilised 
using an extension of the random packing method described above. The spheres are chosen in order of increasing height and, in turn, are allowed to roll and fall into stable positions. In this part of the shake cycle spheres may roll over, and rest on, any other sphere in the assembly. This includes those spheres which are still to be stabilised and which may, 
in turn, undergo further rolls and falls. This is a fully cooperative process, which is crucial for realistic simulations of granular media. Further details of the simulation algorithm, including the use of Gaussian noise to model the random lateral
displacements in the expansion may be found in \cite{MehtaPRL,MehtaPRE}.

In the present investigations, spheres are shaken at 9 amplitudes parametrised in units of sphere diameters: $A = 0.05, 0.08, 0.10, 0.15, 0.18, 0.20, 0.25, 0.28$ and $0.30$. For example, $A = 0.30$ means that spheres are able to move longitudinally by, on average, $0.30$ sphere diameter during a shake cycle. The volume fraction is measured as a function of shaking amplitudes over $10^{5}$ cycles.
We notice that, within a range of excitation amplitudes, there is a sharp increase in packing fraction well above the random close packing density $\phi_{rcp}$. Further shaking for extended periods is seen to produce spontaneous jumps to denser, ordered packings which we have termed \enquote*{spontaneous crystallization}. Our analysis of these packings is divided into two main parts. First, we define  global measures in order to characterize  spatial structures in the system. Second, we define a sphere cluster on which local order metrics are applied, to distinguish between different stages of local cluster development. Our results suggest that the driving force has a critical role to play in the observed competition between hcp and fcc order.

\section{\label{sec:GBOP}GLOBAL ORDER ANALYSIS}

\begin{figure}
\centering
\includegraphics[height=7 cm, width=9 cm]{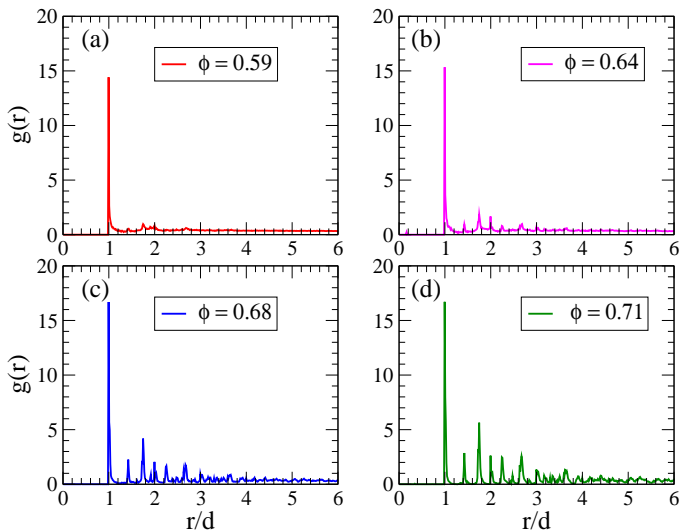}
\caption{(Color online) Plots of radial distribution functions $g(r)$ as a function of normalized distance $r/d$ for various packing fractions $\phi$. The number of peaks shows the development of spatial order from low (Figure~\ref{Figure1} (a)) to high  (Figure~\ref{Figure1} (d)).}
\label{Figure1}
\end{figure}

In this section, we investigate global features of the packings generated by our simulations as a function of shaking amplitude.
The radial distribution function $g(r)$  is the most obvious indicator of order accordingly, we plot it in  Figure~\ref{Figure1} as a function of $r$, for different packing fractions. We note that more and more peaks appear as the packing fraction increases, indicating that spatial ordering has set in. The fact that both fcc and hcp (see for example the peak at $1.91$ in Figure~\ref{Figure1} (c)) peaks are observed already indicates that locally, both types of order are present.

\begin{figure}
\centering
\includegraphics[height=7 cm, width=9 cm]{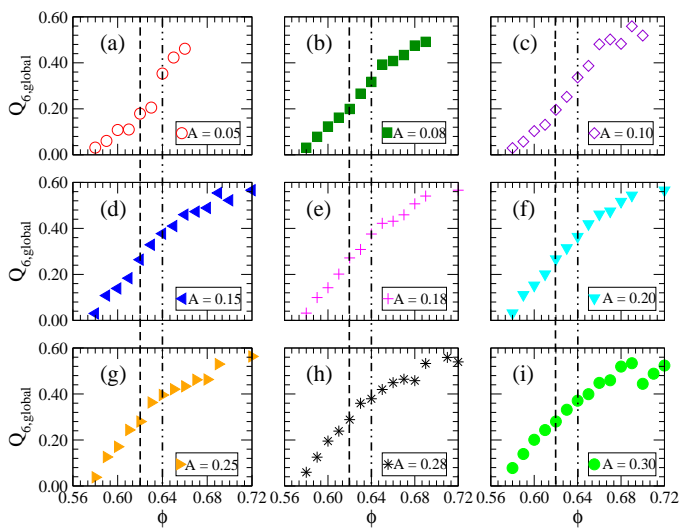}
\caption{(Color online) Variation of $Q_{6,global}$ against packing fraction $\phi$. Note the slight jump (kink) at $0.62$ and steady rise after $0.64$. The global order shows breakdown at 0.69 (Figure~\ref{Figure2} (c), (d), (g), (h) and (i)). The vertical line markers at $ \phi = 0.62$ $(--) $ and 0.64 $(-.-)$ serve as a guide to the eye.}
\label{Figure2}
\end{figure}

In order to do a more detailed analysis, we use the global bond orientation order parameters defined in \cite{Steihardt},

\begin{equation}
\label{Eq1}
Q_{l,global} \equiv \left[\frac{4\pi}{2l+1} \sum_{m=-l}^{l}\Bigl| \langle {{Y_{l}}^m(\Theta(\vec r),\Phi(\vec r))} \rangle \Bigr| ^{2}\right]^{1/2}.
\end{equation}
Here, $Y_{l}^m (\Theta,\Phi) $ are spherical harmonics defined with respect to an arbitrary coordinate system and $l, m$ are integers. The average in Equation \eqref{Eq1} is taken over all the bonds in the system for $100$ configurations, and accordingly  $Q_{6,global}$ is computed for different packing fractions $\phi$ and plotted in Figure~\ref{Figure2} for  the nine amplitudes mentioned above. We mention here that the variation of $Q_{6,global}$ with shaking amplitude is implicit in the figures, since
amplitude governs both the value of the final density $\phi_{max}$ reached in a given time, as well as its rate of change.

A universal feature is that the overall growth of global order towards  $\phi_{max}$ has a kink between
the values of $\phi \sim 0.62$ and $\phi \sim 0.64 $. Recent experiments \cite{Kudroli} on sheared granular spheres suggest that  $\phi \sim 0.62$ is the onset of ordering, while other simulations \cite{Torquato2,Song} suggest that $\phi \sim 0.64 $ is a critical value above which ordered structures are increasingly evident. Our interpretation of results in the context of these facts is that in the kink region, it is likely that the correlation between regions of nucleated ordering increases, until any further increase after  $\phi \sim 0.64 $ leads to appreciable regions of crystallinity; in turn, these become larger as the density is further increased. This is consistent with the interpretation of $\phi \sim 0.64 $ as a critical state.

A feature to note is that the three smallest amplitudes ( $A = 0.05, 0.08, 0.10$) reach lower values of $ \phi_{max}$ (Figure~\ref{Figure2} (a)-(c)) than the rest, which all reach a value of $\phi_{max} \sim 0.72$. This is most likely due to the fact that at smaller amplitudes, the dynamics are much slower, and that perhaps the same maximal densities would be reached for computer times that were inaccessible to us.

Another feature to note is that there appears to be a temporary \enquote*{breaking} of order around $\phi \sim 0.69$ in some cases. This, as well as the kink between  $\phi \sim 0.62$ and $\phi \sim 0.64 $,  motivates a closer examination of local ordering, which will be discussed in the next section.

\section{\label{sec:LBOP}LOCAL ORDER ANALYSIS}

Since the onset of global ordering must have local precursors, we investigate the ordering of local clusters in the rest of this paper. We first define what a cluster means in the present context, since this is the unit on which our local order parameters will be defined. We define a sphere cluster as an assembly of $13$ spheres, as the basic unit of local order. This is motivated by the fact that in stable fcc and hcp structures, a cluster of $12 $ spheres around a central sphere  gives a maximum packing fraction of $\phi = 0.74 $. It is clear from this definition ( previously been used in the structural analysis of colloids \cite{Gasser1} and granular sphere packings \cite{Kudroli,Troadec}) that while sphere clusters are useful for distinguishing
different types of order, they would be the wrong choice for distinguishing order from disorder.

Our main objective in this section is of course to distinguish between fcc and hcp structures -- while methods involving Voronoi diagrams exist \cite{TroadecEPL}, we prefer to use the local bond orientation order parameter due to Steinhardt et al \cite{Steihardt}.
They define the local bond orientation order parameter as
\begin{equation}
\label{Eq2}
Q_{l,local}(i) \equiv \left[\frac{4\pi}{2l+1} \sum_{m=-l}^{l}\left| \sum_{j=1}^{N_{s}(i)}{{Y_{lm}(\Theta_{ji},\Phi_{ji})}}/N_{s}(i)\right|^{2}\right]^{1/2},
\end{equation}
where $Y_{lm}$ are  spherical harmonics, with $ l$ and $ m$ integers. The angles $\Theta_{ji}$ and $\Phi_{ji}$ are polar angles with respect to an arbitrary coordinate system, characterising the bond vector $\vec r_{ij}$ from sphere $j$ to sphere $i$.  The sum and averages in Equation \eqref{Eq2} are computed over all neighbouring spheres $N_{s}(i)$ of sphere $i$. This definition exploits the difference in the stacking sequences of hcp and fcc clusters, for $l = 4, 6$ \cite{Steihardt}. For fcc and hcp sphere clusters the values of the pair ($Q_{6,local}, Q_{4,local}$) are known to be $(0.575, 0.191)$ and  $(0.485, 0.097)$ respectively \cite{Torquato3}.

In our study, we define a nearest neighbour of a sphere as that which lies at a distance of  $1.2$ sphere diameters from it -- this corresponds to the first minimum of the radial distribution function. With the choice of $N_{s}(i) = 12$, we restrict ourselves to spheres which only have $12$ neighbours, corresponding to a sphere cluster as defined above. With these choices,
we compute the local bond orientation order parameters, $Q_{6,local}$ and $Q_{4,local}$ for each sphere with a view to distinguishing between hcp and fcc order.

We divide our local order analysis into three temporal stages with respect to values of density, for each amplitude considered. 
The states corresponding to $\phi \sim 0.61 $ to $\phi \sim 0.65$ are relatively disordered, and we discuss them first. Next, we examine the partial ordering that sets in at $\phi \sim 0.68 $ and $0.69$. Finally, we discuss the most ordered states corresponding to the highest density $\phi_{max}$ achieved for each amplitude. Throughout, we use
scatter plots of $Q_{6,local}$ and $Q_{4,local}$ and non-parametric kernel density plots of $Q_{6,local}$ for displaying our results.

\begin{figure}
\centering
\includegraphics[height=7 cm, width=9 cm]{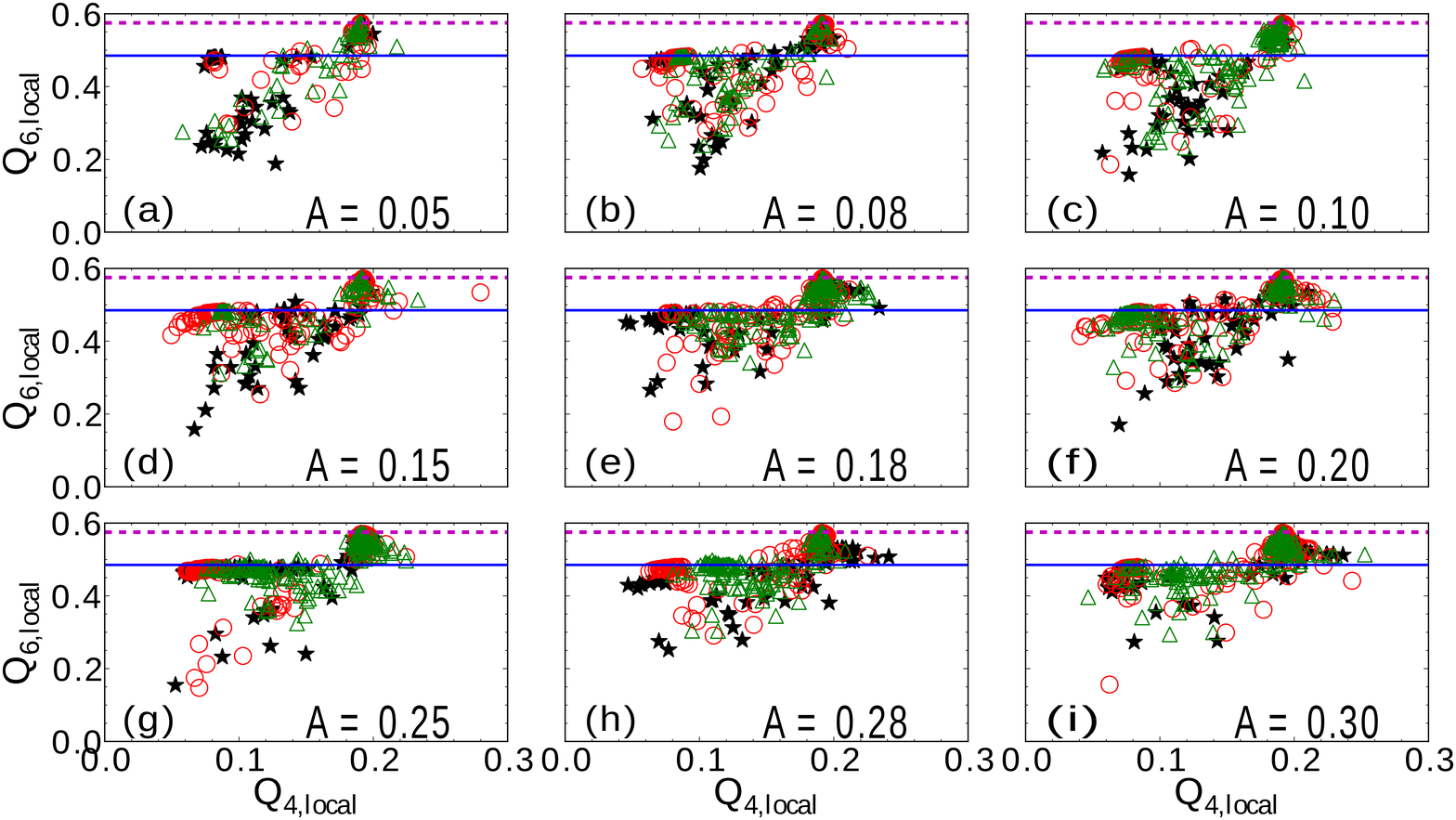}
\caption{(Color online) Plots of $Q_{6,local}$ vs. $Q_{4,local}$. The scattered values show  disordered states of $\phi = 0.61$ (stars), $\phi = 0.62$ (open circles) and $\phi = 0.63$ (open triangles). The horizontal line markers at $0.485$ (blue for hcp) and $0.575$ (magenta for fcc) serve as a guide for the eye.}
\label{Figure3}
\end{figure}

\begin{figure}
\centering
\includegraphics[height=7 cm, width=9 cm]{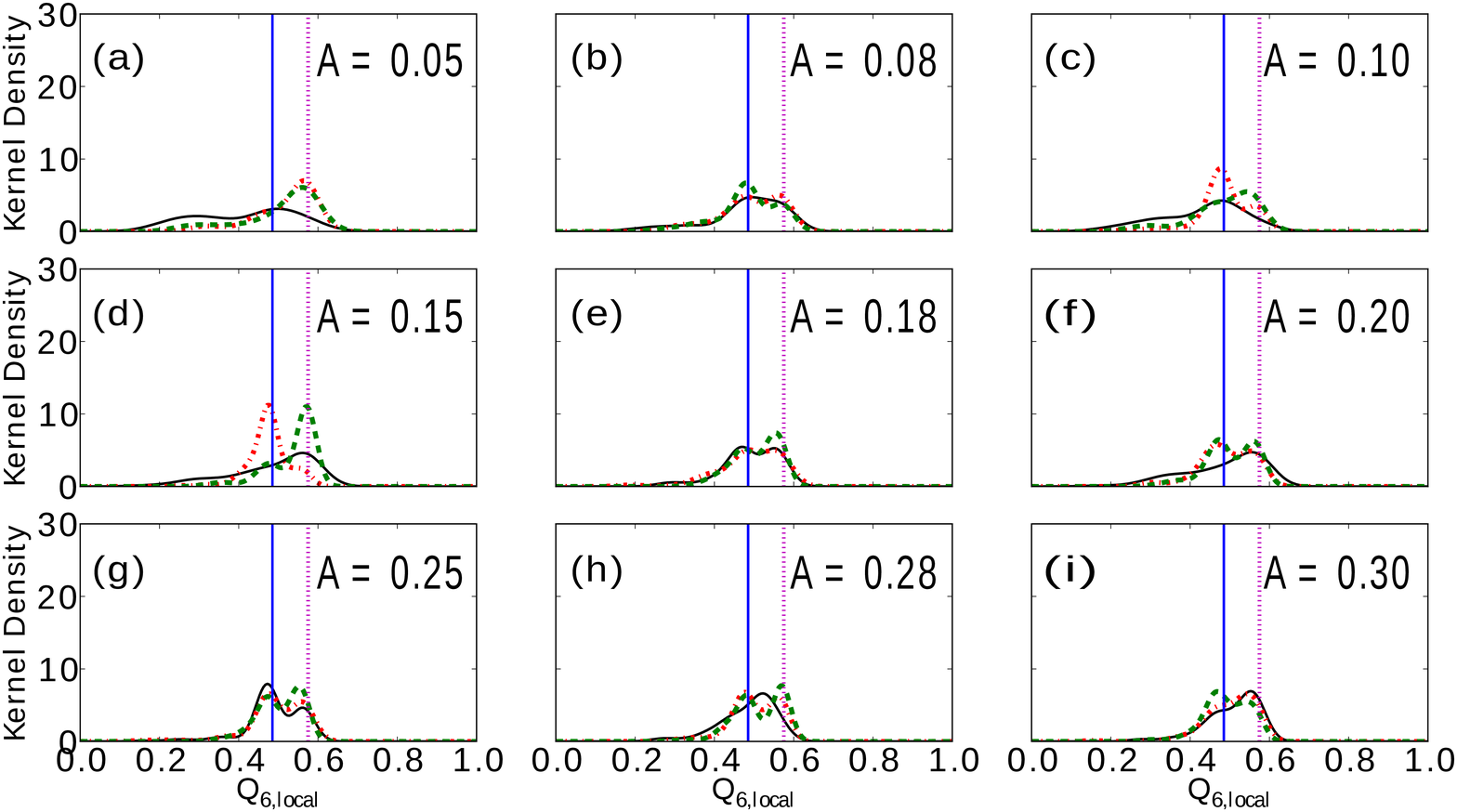}
\caption{(Color online) Probability density plots of a $Q_{6,local}$ for $\phi \sim 0.61$ (black $-$), $0.62$ (red $-.$) and $0.63$ (green $--$). The peaks are broad and robust. The vertical line markers at $0.485$ (blue for hcp) and $0.575$ (magenta for fcc) serve as a guide for the eye.}
\label{Figure4}
\end{figure}

\subsection{\label{sec:LD} DISORDERED SPHERE CLUSTERS AT LOW DENSITIES}
 
The free volume available to a sphere to realign itself with respect to its neighbours is proportional to the shaking amplitude $A$ \cite{am}. Since such collective rearrangement is the catalyst which drives the nucleation of order in a packing, we would expect more rapid nucleation to occur for larger free volumes, i.e. the larger amplitudes in our set of nine. (We note that none of these is of course large enough to cause the assembly to be so fluidised that order never sets in -- for a more detailed discussion of this optimal range of amplitudes, see \cite{MehtaJCP}).

We observe that the number of nucleating sites increases as the density is increased from $\phi \sim 0.61$ to $0.63$. Both the scatter plots of $Q_{6,local}$ vs. $Q_{4,local}$ (Figure~\ref{Figure3}) and the probability density plots of $Q_{6,local}$ (Figure~\ref{Figure4}) confirm that sphere packings in this range of densities are largely disordered at a local level. Note that for $\phi \sim 0.62$ and $0.63$, Figure~\ref{Figure4} shows the onset of double peaked distributions. Both peaks are however relatively broad, indicating that complete crystallization has not occurred in a cluster. This is consistent with our remarks above that $\phi \sim 0.62$ could possibly be thought of as the onset of crystallisation.

\begin{figure}
\centering
\includegraphics[height=7 cm, width=9 cm]{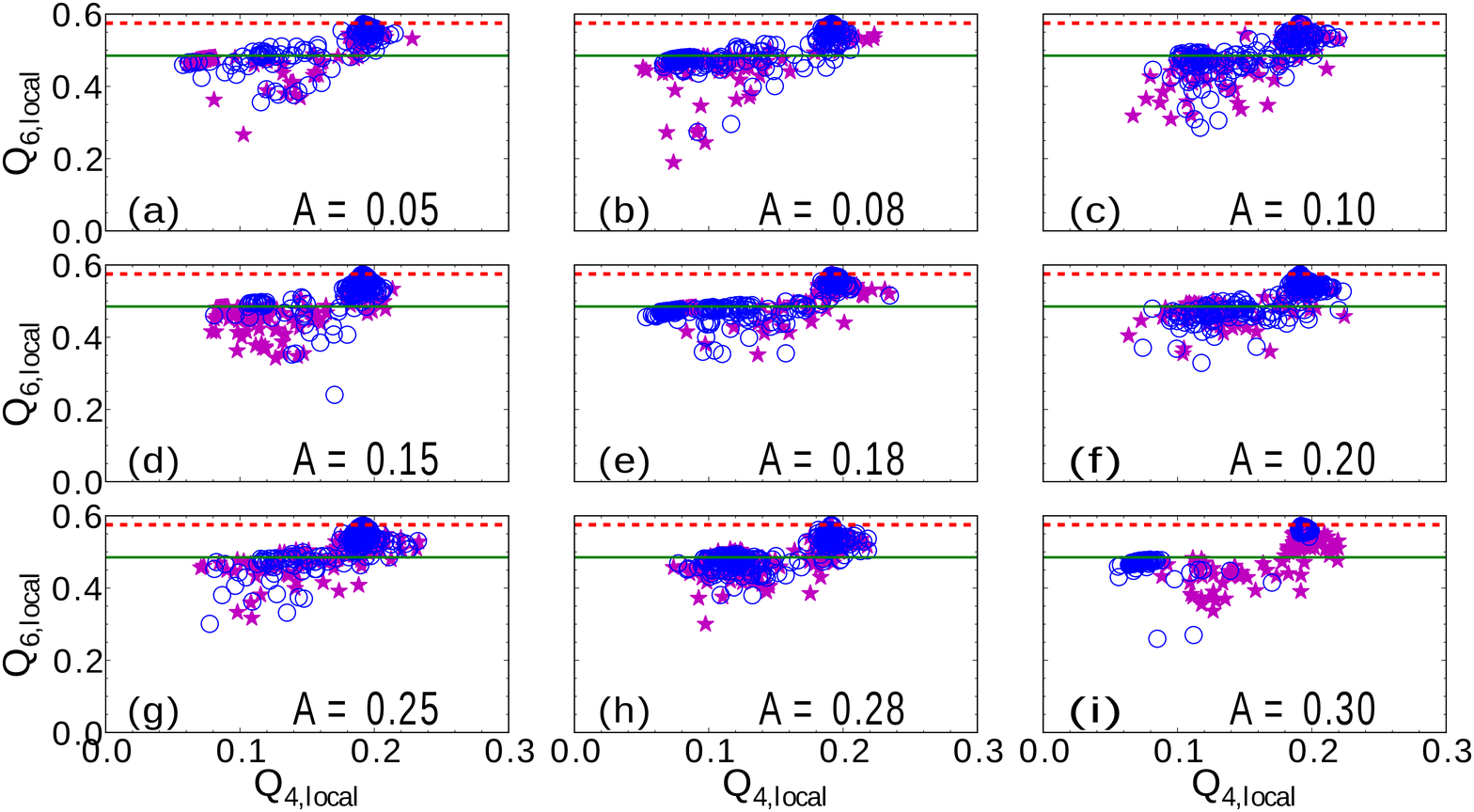}
\caption{(Color online) Plots of $Q_{6,local}$ vs. $Q_{4,local}$ for the densities $\phi\sim 0.64$ (stars) and $0.65$ (open circles). Both the states are disordered. The $\phi\sim 0.65$ state has less scatter than $0.64$. The horizontal line markers at $0.485$ (green for hcp) and $0.575$ (red for fcc) serve as a guide for the eye.}
\label{Figure5}
\end{figure}

\begin{figure}
\centering
\includegraphics[height=7 cm, width=9 cm]{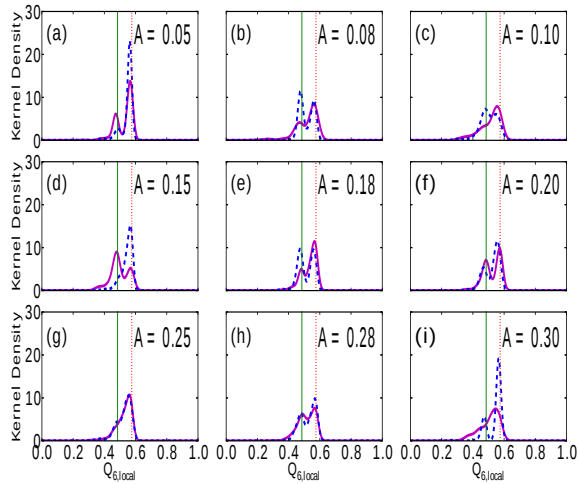}
\caption{(Color online) Probability density plots of a $Q_{6,local}$ for $\phi \sim 0.64$ (magenta $-$), $0.65$ (blue $--$). The distributions have sharper peaks than before with some predominance of the second peak. The vertical line markers at $0.485$ (green for hcp) and $0.575$ (red for fcc) serve as a guide for the eye.}
\label{Figure6}
\end{figure}

Partially ordered sphere clusters begin to make their presence felt at $\phi \sim 0.64$ and $0.65$ respectively. These are less disordered  (Figure~\ref{Figure5}) than those at lower densities. We notice that there is less scatter at the four highest densities than in the rest, indicating, as mentioned above, that ordering has been facilitated by access to greater free volume.
An examination of Figure~\ref{Figure6} shows sharper peaks overall compared to Figure~\ref{Figure4}, indicating a greater proportion of ordered sphere clusters. The second peak of $Q_{6,local}$ densities is more consistently observed than the first peak in Figure~\ref{Figure6}, indicating a preponderance of fcc ordering.


\subsection{\label{sec:POS} NEARLY ORDERED PACKINGS AT HIGHER DENSITIES: THE COMPETITION BETWEEN FCC AND HCP}

At higher densities corresponding to $\phi \sim 0.68 $ and $0.69$, order sets in increasingly. This is accomplished both
by an increase in the number of sphere clusters as well as the degree of ordering within each one. Consequently, the important issue is the competition between hcp and fcc ordering (rather than the competition between order and disorder). We notice accordingly that there is now a tendency for the sphere clusters to cluster around the fcc and hcp values, a process which is much sharper for the higher of the two densities. In this case, for $\phi \sim 0.69 $, there is an interesting phenomenon at  $A = 0.10$ and $A= 0.15$, when it seems that the sphere clusters are entirely characterised by fcc ordering, while both fcc and hcp ordering are back in play at higher amplitudes. The probability density plots of Figure~\ref{Figure8} reinforce these claims, as expected: also, as expected, the peak probability densities for $\phi\sim 0.69$ are sharper than those for $0.68$. 

We speculate that the dearth of free volume at $A = 0.10$ and $A= 0.15$ could have led to the interruption of the evolution into hcp ordering, since this seems to set in for higher amplitudes. Although more needs to be done to verify this, it is tempting to think that there might well be an optimal range of amplitudes (not so low that the ordering process is incomplete, not so high as to allow the free evolution into competing structures) at these high densities where spontaneous crystallisation into a pure fcc state might occur.

Finally, if we recall that a breakdown of global ordering was observed at $\phi \sim 0.69 $, our local ordering analysis suggests that the interfaces between crystallites of fcc and hcp might be responsible for this.

\begin{figure}
\centering
\includegraphics[height=7 cm, width=10 cm]{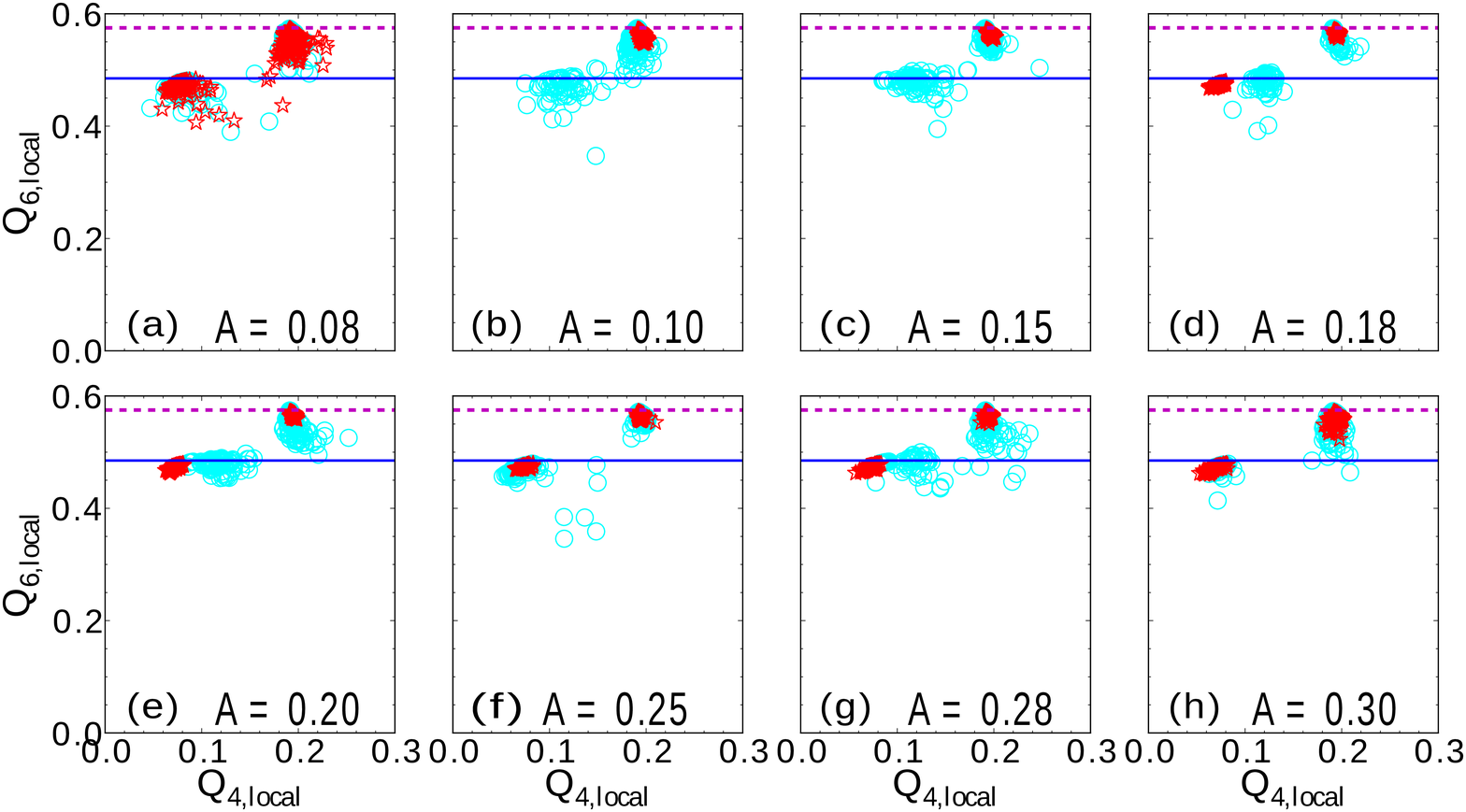}
\caption{(Color online) Plots of $Q_{6,local}$ vs. $Q_{4,local}$ for densities $\phi \sim 0.68$ (open circles) and $0.69$ (stars). The state of $\phi\sim 0.68$ has more scatter than $0.69$. Notice the sharp division into two distinct groups for $\phi \sim 0.69$. The special noticeable situations are for $A = 0.10$ (Figure~\ref{Figure7} (b)) and 0.15 (Figure~\ref{Figure9} (c)). The horizontal line markers at $0.485$ (blue for hcp) and $0.575$ (magenta for fcc) serve as a guide for the eye.}
\label{Figure7}
\end{figure}
\begin{figure}
\centering
\includegraphics[height=7 cm, width=10 cm]{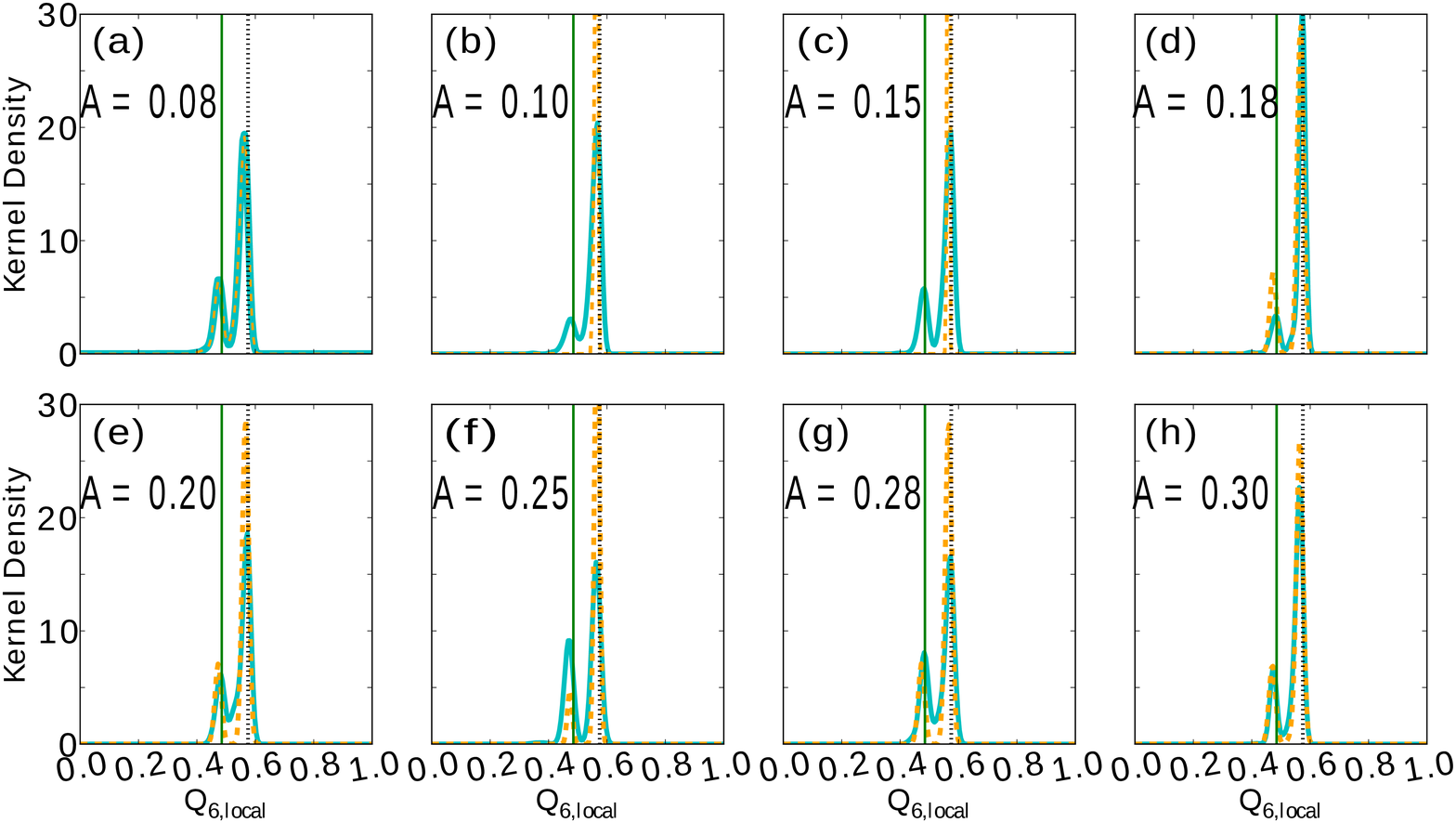}
\caption{(Color online) Probability density plots of  $Q_{6,local}$ for $\phi \sim 0.68$ (cyan $-$), $0.69$ (orange $--$). The peaks of $\phi \sim 0.69$ are sharper than those of  0.68. For $\phi \sim 0.69$ at $A = 0.10$ and 0.15 only one sharp peak at $Q_{6,local} \sim 0.575$ is visible (Figure~\ref{Figure8} (b) and (c))). The vertical line markers at $0.485$ (green for hcp) and $0.575$ (black for fcc) serve as a guide for the eye.}
\label{Figure8}
\end{figure}

\subsection{\label{sec:CMD} ORDERING AT ASYMPTOTIC DENSITIES}

In this section, we see a stronger illustration of some of the ideas floated in the earlier subsection, as we examine the ordering that sets in at the highest densities  $(\phi_{max})$  achieved for each of the amplitudes considered. For the lowest amplitudes,
the ordering process is clearly incomplete, and for the highest amplitudes, hcp and fcc ordering coexist. An intermediate, \enquote*{optimal} range of amplitudes where single crystals of fcc emerge, is also observed.

\begin{figure}
\centering
\includegraphics[height=7 cm, width=9 cm]{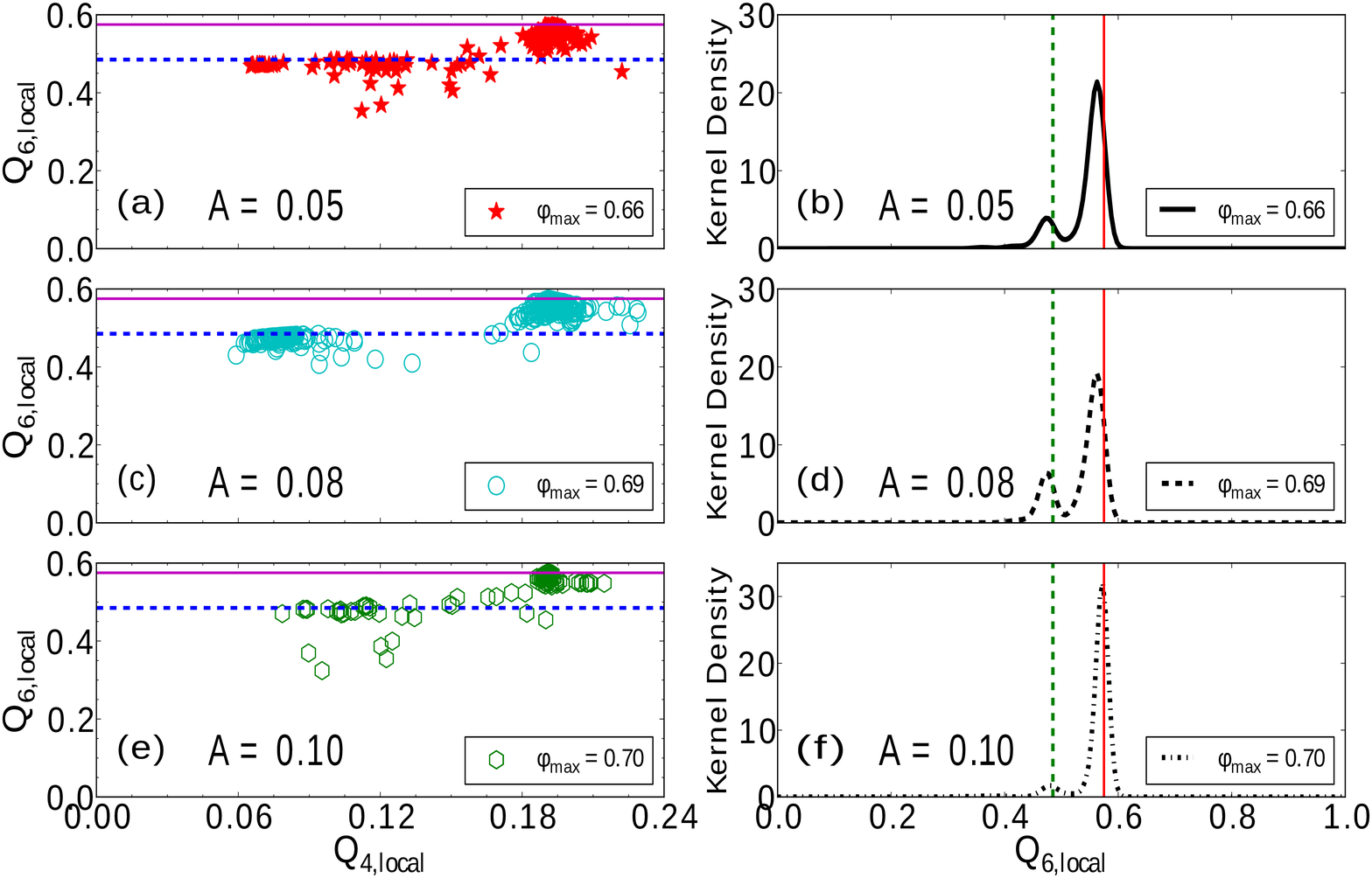}
\caption{(Color online) Plots of the maximum densities for $A = 0.05$, $0.08$, and $0.10$. Scatter plots of $Q_{6,local}$ vs. $Q_{4,local}$ are broadly distributed into two groups. The probability density plots (Figure~\ref{Figure9} (b), (d) and (f)) indicate a second peak which is sharper than the first. The horizontal and vertical lines at $0.485$ (for hcp) and $0.575$ (for fcc) serve as a guide for the eye.}  
\label{Figure9}
\end{figure}

For the three lowest amplitudes $A = 0.05$, $0.08$ and $0.1$, we observe clearly that for the computer times at our disposal, full crystallisation did not occur. Figure~\ref{Figure9} shows the scatter plots and the probability density plots for each case. In every case, $Q_{6,local}$ the scatter plots are weighted around the fcc value of 0.575. That fcc ordering is predominant is more clearly reflected in the sharp second peak of $Q_{6,local}$ (Figure~\ref{Figure9} (b), (d), (f)). Of course we cannot rule out a further evolution when the system is shaken for longer times and in fact we would expect more complete ordering to emerge
in that limit, even for the lowest shaking amplitudes.

\begin{figure}
\centering
\includegraphics[height=7 cm, width=9 cm]{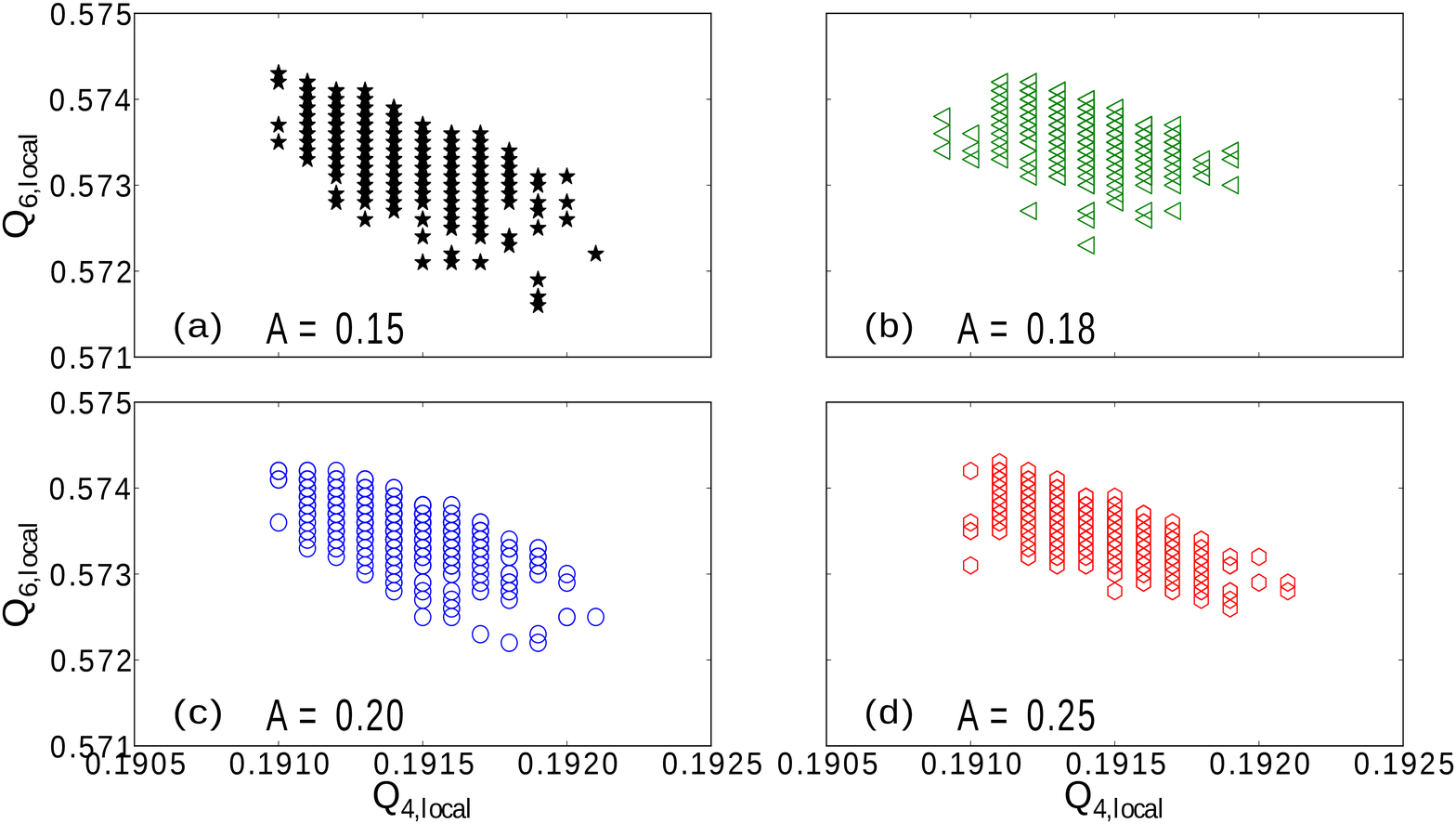}
\caption{(Color online) The scatter plots of $Q_{6,local}$ vs. $Q_{4,local}$ show a single fcc cluster for a maximum density of $\phi \sim 0.72$ for $A = 0.15$ $0.18$, $0.20$, and $0.25$.}
\label{Figure10}
\end{figure}
\begin{figure}
\centering
\includegraphics[height=7 cm, width=9 cm]{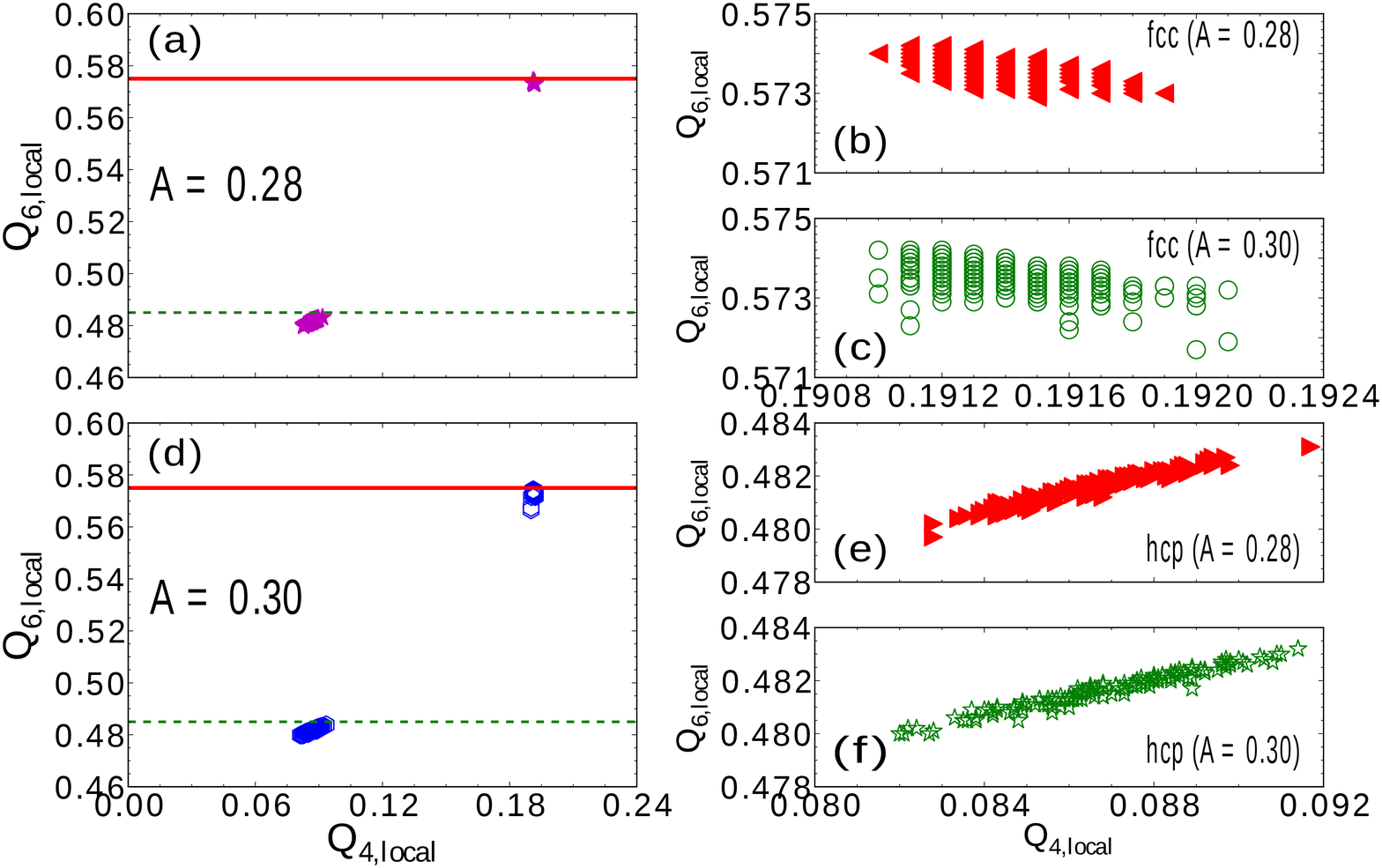}
\caption{(Color online) Graphs of $Q_{6,local}$ vs. $Q_{4,local}$ show the coexistence of two fcc and hcp sphere-clusters. The data of fcc and hcp cluster is  again plotted in Figures~\ref{Figure11} (c), (d) and (e), (f) for the respective amplitudes. The horizontal line markers at $0.485$ (green for hcp) and $0.575$ (red for fcc) in Figure~\ref{Figure11} (a) and (b) serve as a guide for the eye.}
\label{Figure11}
\end{figure}

For intermediate amplitudes ($A = 0.15$, $0.18$, $0.20, 0.25$), a single fcc phase appears at the asymptotic density $\phi_{max} \sim 0.72$ (Figure~\ref{Figure10}). This seems very robust, lending weight to our arguments that an optimal range
of amplitudes exists for spontaneous crystallisation into a single fcc state. It would be interesting if this phenomenon could be probed experimentally as well as by independent simulations, both from the point of view of theory as well as application.

For higher amplitudes still, we see a clear separation of the two kinds of ordering, centred on the lines corresponding to the fcc and hcp values (Figure~\ref{Figure11}). This coexistence reinforces the conclusions of previous simulations \cite{Troadec, Makse, Dong}. However, fcc ordering still predominates: the relative fraction of fcc sphere clusters, given by $N_{fcc}/(N_{fcc} + N_{hcp})$ is $0.78$ for $A = 0.28$ and $0.77$ for $0.30$, where $N_{fcc}$ and $N_{hcp}$ are the numbers of fcc and hcp sphere clusters respectively.

Our main conclusion is therefore that there is full ordering at the final densities corresponding to the highest amplitudes in our list, but that in all probability there are the analogue of dislocations which separate regions of hcc and fcc ordering. These dislocations would represent the deviation from the perfect global ordering that obtains for the intermediate amplitudes.

We emphasize of course that these results are valid for the time of shaking we have considered, and so cannot rule out further crossovers at larger times. Although studies have investigated the coexistence of these two cluster types in colloids and granular materials as a function of shear rate, we believe that this is the first attempt to analyse crystalline clusters systematically by varying shaking amplitudes.

\section{\label{sec:CONL} CONCLUSIONS}

We have carried out computer simulations of shaken granular packings over a range of amplitudes. The highest amplitudes we chose were still well within the range where collective motion predominates \cite{MehtaPRL}, i.e. those where there is insufficient free volume for most spheres to move independently of each other. We have observed that spontaneous crystallisation occurs in our chosen dynamical regime, in the limit of long vibration times. Our observations of global order show that there is a region of increasing nucleation between the onset of order and the random close packed limit, which deserves further investigation. Also,
we noted an apparent breakdown of global order at higher densities, which our local order parameters suggest may be due to interfaces between crystallites of fcc and hcp. Our observations of local order also suggest that at the highest packing densities, there may be an optimal range of amplitudes where crystallisation into a single fcc state occurs. Amplitudes even higher than this lead to a coexistence of hcp and fcc order, with the latter predominating; we suggest that dislocations between the two sorts of ordering should be observed, and hope that further work to investigate this important issue will be undertaken.

\section{ACKNOWLEDGEMENTS}
DPS would like to thank CSIR, India, for financial support.

\end{document}